\documentclass[10pt]{article}

\usepackage{amsmath}
\usepackage{amssymb}
\usepackage{amsthm}
\usepackage{graphicx}
\usepackage{algorithmic}

\usepackage{cite}

\usepackage{color} 

\usepackage[labelfont=bf,labelsep=period,justification=raggedright]{caption}

\makeatletter
\renewcommand{\@biblabel}[1]{\quad#1.}
\makeatother

\date{}

\begin{document}

\begin{flushleft}
{\Large
\textbf{Transforming a random graph drawing into a Lombardi drawing}
}
\\
Nicolaos Matsakis$^{1}$
\\
\bf{1} Department of Information and Computer Science, UC Irvine, CA, USA
\\
$\ast$ E-mail: nmatsaki@uci.edu
\end{flushleft}

\section*{Abstract}

The visualization of any graph plays important role in various aspects, such as graph drawing software. Complex systems (like large databases or networks) that have a graph structure should be properly visualized in order to avoid obfuscation. One way to provide an aesthetic improvement to a graph visualization is to apply a force-directed drawing algorithm to it. This method, that emerged in the 60's views graphs as spring systems that exert forces (repulsive or attractive) to the nodes. The first force-directed drawing algorithm of Tutte \cite{Tut63} was aiming at providing a barycentric representation of the graph and it is suitable  for getting a straight-line drawing of 3-connected planar graphs without crossings. Each vertex is gradually placed at the berycenter of its neighbour vertices. This method used no spring systems, though it can be easily understood that springs can be applied to the system getting the same result as that of the barycentric method. Unfortunately, Tutte's method usually gives bad angular resolutions.

Many other force-directed methods have appeared since then. Eades' method \cite{eades} creates 2D layouts using spring systems and it is suitable for graphs with less than 30 nodes. Attractive forces are applied to adjacent vertices only and repulsive to all pairs of vertices (so we can view the edges as springs with zero rest length and the nodes as electrically charged identical particles). Of course, the exact type of the forces exerted is a very important aesthetic criterion. Eades sets logarithmic attractive forces and inverse square root repulsive forces. Fruchterman and Reingold \cite{DBLP:journals/spe/FruchtermanR91} apply the notion of simulated annealing, in order to provide an even better 2D layout of less cumulative spring energy, compared to that of Eades. The main disadvantage is, again, that the algorithm is limited to graphs having less than 40 vertices. Hadany and Harel \cite{HH99} provide a force-directed layout for much larger graphs of over 1000 vertices. In order to obtain that, they follow a multi-scale approach where they initially consider an abstraction of the graph, which is iteratively augmented by the graph details. Gajer, Goodrich and Kobourov \cite{DBLP:journals/comgeo/GajerGK04} use a coarsening strategy which avoids quadratic space and time complexities. They use a vertex filtration which enables the algorithm to restrict the number of vertices relocated. They, also, introduce the idea of realizing the graph in high-dimensional Euclidean space. An excellent survey on various force-directed methods can be found in \cite{tamassia-book}.

A Lombardi drawing of a graph is a drawing where the edges are drawn as circular arcs (straight edges are considered degenerate circular ones) with perfect angular resolution. This means, that consecutive edges around a vertex are equally spaced around it. In other words, each angle\footnote{ every angle in this survey will be expressed in rads} between the tangents of two consecutive edges is equal to $2\pi/d$ where d is the degree of that specific vertex. This kind of drawing took its name from the American artist Mark Lombardi, who mainly used this technique in his paintings. The requirement of using circular edges in graphs when we want to provide perfect angular resolution is necessary, since even cycle graphs cannot be drawn with straight edges when perfect angular resolution is needed.

In this survey, we provide an algorithm that takes as input a random drawing of a graph and provides its Lombardi drawing. 

\section{Lombardi Graphs}

We will use the spring-force layout of a graph. A very important property of a single vertex which is connected to a number of d vertices by circular edges is that, if the edges are regarded as circular springs, the total spring force exterted on the central vertex will be equal to the one exerted on it if the circular springs are replaced by straight ones, each one having the same length as the respective circular arc and a direction onto the tangent of each respective circular edge (as in figure 1). The reason behind this, is that each force is exerted instantly on the direction of each tangent. In figure 1 there is an example of a node having degree equal to 4. Whether we take the 4 circular edges or the 4 straight ones (of equal corresponding length) under account, the instant result on the move of the central vertex will be the same.

\begin{figure}
\begin{center}
\includegraphics[scale=0.30,trim = 25mm 40mm 20mm 5mm, clip]{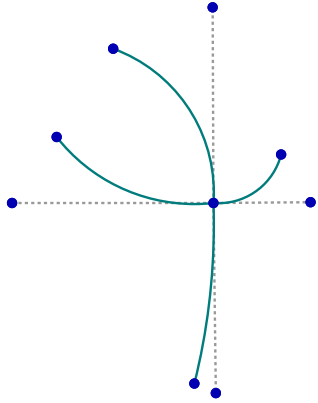}
\caption{The spring forces on the central vertex due to the connection with the 4 circular edges-springs, form a cumulative force equal to the sum of the spring forces of 4 straight spring-edges (dotted lines), each one having the same length as that of its respective circular arc. In the specific example, however, we can easily see that the cumulative force on the central vertex is not zero, something which means that our algorithm will move the central vertex to a position where $\sum_{i=1}^{4}\stackrel{\rightarrow}{F_{i}}=0$}
\end{center}
\end{figure}

We will use this relation extensively while forming the angular resolution around a vertex. Our procedure actually uses, initially, an already known force-directed method and, afterwards, starts fixing the angular resolution around each vertex of the current drawing. The target is to obtain a drawing with the least cumulative spring energy, which will, also, have perfect angular resolution.

The type of force-directed method initially applied, depends on the number of nodes of G. Let's assume that this is not great enough (at most 40 vertices), so we can apply the barycentric method of Tutte. However, there are methods, as predescribed, for much greater numbers of nodes. After applying the barycentric method, we order the nodes of the graph according to a descending order of degree. We start by picking the first vertex in the order and choosing its neighbour vertex that is farthest away in straight distance from it\footnote{the reason behind this is that the minimum total energy that we want to achieve forces us to connect initially the 2 farthest points in plane with a straight edge; any other edge gives greater length and this means greater energy} (for example we pick vertex b in figure 2). We erase the current edge (circular or straight) between them and connect them with a straight edge (so in a case of a previous straight edge connection, we do not change anything). We continue in a clockwise order to the next adjacent edge (that is c in the example), which we erase and connect the 2 corresponding vertices (the central and its neighbour vertex) with a circular edge which has angle $2\pi/d$ with the previously formed edge (in anticlockwise order and we obviously refer to angles between tangents in the position of the central vertex). This is totally valid since any two points in plane can be connected with one circular arc, if we specify a tangent passing through one of them. We continue doing this procedure, creating equal angles between tangents, until we reach the very first edge, where we stop.

\begin{figure}
\begin{center}
\includegraphics[scale=0.30,trim = 25mm 10mm 20mm 5mm, clip]{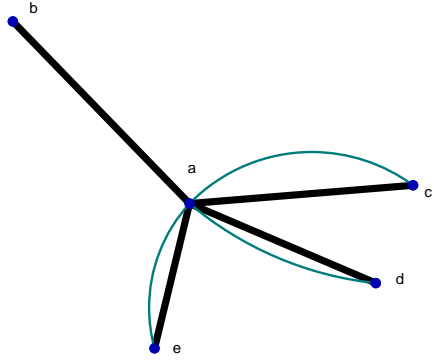}
\caption{The central vertex a has degree 4, so starting with neighbour vertex b and continuing in clockwise order, we arrange each new circular edge having angle equal to $\pi/2$ with the exactly previous one, until we reach again b where we stop. Bold (straight) edges refer to the previous edge visualization.}
\end{center}
\end{figure}

It is obvious, now, that the Lombardi property has been applied on the central examined vertex. However, the zero cumulative spring force has been lost because the spring forces have changed, in both the direction and the length of the springs. So, we must fix that, by finding the expression of a spring force on a vertex when the spring is circular. Here is where the first observation is used. We must bring the central vertex to the position of zero cumulative spring force, leaving stable its neighbour vertices. So, our target is to minimize the energy of the spring system of this examined central vertex which is connected to d vertices under the invariance of the Lombardi property, which means that we are not allowed to alter the angles between the tangents of the edges and the x-axis, formed in the plane. In other words we seek to minimize the sum of the circular springs' lengths which equals $\sum_{i=1}^{d}t_{i}$, where $t_{i}$ is the arc length of the i-th adjacent edge (starting from the one we beginned with in the previous step and continuing in clockwise order\footnote{so in our example $t_{1}$ is the circular edge (straight edges are regarded as circular arcs of infinite radius) connecting a with vertex b}). When we accomplice that and move the position of the central vertex to a position of mimimum total spring energy, we continue with the rest of the vertices in the descending order and then we repeat the whole algorithm from the beginning several times to gain convergence. The number of times may vary for each drawing, so we can just calculate the difference in the co-ordinates of every vertex between each 2 repetitions of the algorithm and stop where we have reached the desired accuracy, for each one of the vertices.

\section {The Algorithm}

A basic trigonometric relation is that the length of a chord x is equal to \begin{equation} x=t*sin(\theta)/\theta \end{equation} where t is the length of the arc that is defined between the 2 points of the chord and $\pi/2\geq\theta>0$ is the angle between the chord x and the tangent at any of the 2 intersection points of the defined cycle and the chord, as we can see in Figure 3.

\begin{figure}
\begin{center}
\includegraphics[scale=0.27,trim = 25mm 60mm 20mm 50mm, clip]{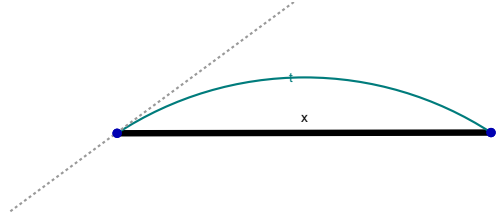}
\end{center}
\caption{For any given arc length t, the corresponding chord length obeys the relation $x=t*sin(\theta)/\theta$ where $\theta>0$ is the angle between x and the tangent of the arc in the intersection point between the chord and the arc.}
\end{figure}

The Euclidean distance between any points $(p_{1},p_{2})$ in plane (where $p_{1}=(x_{1},y_{1})$, $p_{2}=(x_{2},y_{2})$) is given by the relation \begin{equation} D(p_{1},p_{2})=\sqrt{(x_{1}-x_{2})^{2}+(y_{1}-y_{2})^{2}} \end{equation}

Since we seek minimization of the sum of arc lengths, while keeping the Lombardi property, we want to minimize the quantity \begin{equation} \sum_{i=1}^{d}x_{i}*\theta_{i}/sin(\theta_{i})  \end{equation} where $x_{i}$ refers to the distance between the central vertex and the i-th neighbour vertex (assuming a clockwise enumeration as indicated before from $i=1$ to d) and $\theta_{i}$ is the angle between the tangent of a circular edge connecting the central vertex with the i-th vertex neighbour of it and the straight distance between the central vertex and the currently examined neighbour of it $(0<\theta\leq\pi/2$). In order to avoid obfuscation with the co-ordinates we will refer to the distance $x_{i}$ as D from now on, associated with each specific central examined vertex.
\\

From (2),(3) we get:

\begin{equation} min_{x_{0},y_{0}}\sum_{i=1}^{d}\theta_{i}*\sqrt{(x_{i}-x_{o})^{2}+(y_{i}-y_{o})^{2}}/sin(\theta_{i}) \end{equation}

The Lombardi property of equally spaced tangents on arc edges must be preserved while moving the central vertex on plane. In other words, we move the central vertex trying to obtain $\sum_{i=1}^{d}\stackrel{\rightarrow}{F_{i}}=0$ where $\stackrel{\rightarrow}{F}_{i}$ refers to the spring force exerted between the central vertex and the i-th neighbour vertex of it, assuming zero rest length on each spring.

\begin{figure}
\begin{center}
\includegraphics[scale=0.21,trim = 10mm 30mm 20mm 10mm, clip]{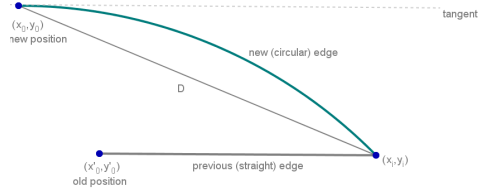}
\end{center}
\caption{The central vertex is moved around trying to minimize the total cimulative spring force exterted on it and preserving the Lombardi property. Here we can see the central vertex moving to a new position along with one of its neighbours which stays stable. Of course the previous edge may have also been circular and not straight as here, which means that we should have used its tangent.}
\end{figure}

However, $sin(\theta_{i})$ is equal to the distance between the i-th neighbour node's (stable) position and the tangent of the specific arc (see figure 4) divided by the new Euclidean distance between the 2 vertices (new position of central vertex and the stable position of its neighbour vertex) which is equal to ${\sqrt{(x_{i}-x_{o})^{2}+(y_{i}-y_{o})^{2}}}$, where the co-ordinates $(x_{o},y_{o})$ refer to the new position of the central vertex. Assume that the central node's initial position is $(x^{'}_{0},y^{'}_{0})$. In order to avoid complicated equations we think of the tangent of the first formed edge as tantamount to the semiaxis \textit{Ox} and since we preserve the Lombardi property while moving around the central vertex, all the tangents keep the same ordering in plane (e.g. each one moves in parallel to its previous arrangement). This fact helps us constructing the arrangement of the d tangents if we set the first one onto semiaxis \textit{Ox} where we think of the new position of the central vertex on (0,0). Therefore, we obtain $y=0$ for the first tangent\{footnote{which we call initial from now on, refering to each specific central examined vertex}, $y=tan(-(i-1)*2\pi/d)x=tan(-2*\pi/d)x$ for the second one (by setting $i=2$) and so on, until we get to the d-th tangent. 

The only thing remaining is to identify the distance between each tangent and the corresponding neighbour vertex. In general, the distance between a straight line (identified by 2 points) and a point in plane, is given by the following equation,  where the co-ordinates refer to figure 5:

\begin{equation}
D=\frac{(x_{b}-x_{0})(y_{0}-y_{a})-(x_{0}-x_{a})(y_{b}-y_{0})}{\sqrt{(x_{b}-x_{0})^{2}+(y_{b}-y_{0})^{2}}}
\end{equation}

\begin{figure}
\begin{center}
\includegraphics[scale=0.27,trim = 21mm 40mm 20mm 32mm, clip]{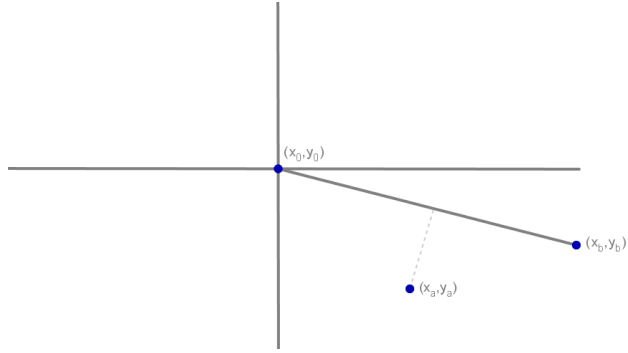}
\end{center}
\caption{The distance between point $(x_{a},y_{a})$ and to the straight line defined on the line segment connecting points $(x_{0},y_{0})$ and $(x_{b},y_{b})$.}
\end{figure}

In our case and since we define each tangent on a semiaxis starting at point (0,0) we surely know point $(x_{0},y_{0})$. Point $(x_{a},y_{a})$ is the actual neighbour vertex i correlated with the specific tangent and point $(x_{b},y_{b})$ can be obtained by setting $x_{b}=1$ and using the tangent form $(y_{b}/x_{b})=tan(-(i-1)*2\pi/d)$ for the specific $i$, to obtain $y_{b}$. The overall distance between the central node's new position and the i-th neighbour node's (stable) position is, then:

\begin{equation}
D=\frac{|-y_{i}+x_{i}*tan(-(i-1)*2\pi/d)|}{\sqrt{1+tan^{2}((i-1)2\pi/d)}}
\end{equation}

Using (6), we get for the $sin(\theta_{i})$:

\begin{equation}
sin(\theta_{i})=\frac{D}{\sqrt{(x_{i}-x_{0})^{2}+(y_{i}-y_{0})^{2}}}
\end{equation}

Replacing (7) into (4), we take a relation solely depending on the central node's and i-th neighbour's coordinates. The sum of it, according to all neighbour vertices, can be minimized, moving the central vertex to a position of minimum total energy:

\begin{equation}
min_{x_{0},y_{0}}\sum_{i=1}^{d}\frac{D*arcsin(\frac{D}{\sqrt{(x_{i}-x_{0})^{2}+(y_{i}-y_{0})^{2}}})}{\frac{D}{\sqrt{(x_{i}-x_{0})^{2}+(y_{i}-y_{0})^{2}}}}
\end{equation}

which is equal to 

\begin{equation}
min_{x_{0},y_{0}}\sum_{i=1}^{d} [(x_{i}-x_{0})^{2}+(y_{i}-y_{0})^{2}]*arcsin(\frac{D}{\sqrt{(x_{i}-x_{0})^{2}+(y_{i}-y_{0})^{2}}})
\end{equation}

We note that D is defined in (6) and that there is no problem with the range of values of the arcsine function since $0<\theta\leq \pi/2$.



After altering the position of the central vertex and moving it to a balancing position, we apply the exact same procedure on the next vertex of the descending order and so on until we reach the $|V|$-th vertex in the order. There, we have to repeat the whole procedure from the beginning since the positions of the vertices should have changed, as many times as it needs for convergence under a desired accuracy.

Finally, we must note that each repetition of the algorithm on a different central vertex should be coming along a rotation of the Cartesian coordinates in order to place the straight distance between the old position of the central vertex and the farthest edge onto semiaxis 0x, as we have described before. So, each time we switch to another central vertex in the descending order we should place the first tangent onto the new 0x semiaxis by multiplying the coefficients of all tangents (the first point needed to describe them is always (0,0) regarding all of them which is not rotated obviously, so we only apply the transformation on each second point) by 

\[ \left( \begin{array}{cc}
cos(\phi) & -sin(\phi)  \\
sin(\phi) & cos(\phi) \end{array} \right)\] 
which corresponds to the clockwise rotation matrix of the 2-dimensional Cartesian coordinates, where angle $\phi$ is the angle between the 2 semiaxes (intersected at (0,0) of the two initial tangents of any 2 consecutive examined central vertices).

The pseudocode follows:
\\

\begin{algorithmic}
\REPEAT

\STATE Apply a barycentric method to G=(V,E)
\STATE Arrange the vertices of the graph according to descending order of degree;
\STATE Let $v=|V|$ and $v_{i}$ be the ordering;
\STATE //start fixing tangents
\STATE //there are no tangents in the beginning; these are formed in the first iteration of the FOR procedure 
\STATE //and their direction stays unaltered unless we unmark them.

\FOR {$i=1$ \TO v}
   \STATE Find the neighbour vertex of $v_{i}$ which is farthest away in straight distance from it; 
   \STATE Delete current edge between them and connect them with a straight edge, marking it with a tangent intersecting the central vertex;
   \STATE Change (Rotate) Cartesian co-ordinates in order to place this (initial) tangent onto semiaxis Ox;
   \STATE Get the next edge of the central vertex in clockwise order;
   \STATE start applying the Lombardi property around the central vertex
   \FOR {$j=1$ \TO d-1}
        \STATE Delete current edge between $v_{i}$ and $j$ and connect them with a circular edge having angle $2j\pi/d$ with the initial tangent;
        \STATE Mark the tangent on this circular edge intersecting the central vertex;
        \STATE $j\gets j+1$
   \ENDFOR
   \STATE Move the central vertex to a position of zero cumulative force on it according to equation (8) by not altering the directions of marked tangents in plane;
   \STATE When the position of the central vertex gives zero cumulative force according to (8), set $i\gets i+1$ and unmark all tangents;
\ENDFOR
\UNTIL {the difference on the co-ordinates of each vertex converges}
\end{algorithmic}

\section{Conclusions}

We have sketched an algorithm for creating a Lombardi drawing of a graph, given a random drawing of the graph. However, the algorithm's convergence remains to be tested for various numbers of graph nodes, since force-directed methods, in general, experience different convergence attitudes, which are highly correlated with the number of nodes of the graphs they are applied on. Especially in our case, graphs with very high degrees on vertices should be checked for appropriate convergence.

\bibliography{sigproc}		
\end{document}